\documentclass{PoS}

\title{A systematic study of the strong interaction with ${\bar{\rm P}}$ANDA}

\ShortTitle{Strong interaction studies with ${\bar{\rm P}}$ANDA}

\author{\speaker{Johan Messchendorp} for the ${\bar{\rm P}}$ANDA collaboration\\
        KVI, University of Groningen, The Netherlands\\
        E-mail: \email{messchendorp@kvi.nl}}

\abstract{The theory of Quantum Chromo Dynamics (QCD) 
reproduces the strong interaction at distances much shorter 
than the size of the nucleon. At larger distance 
scales, the generation of hadron masses and confinement
cannot yet be derived from first principles on basis of QCD.
The ${\bar{\rm P}}$ANDA experiment at FAIR will address the 
origin of these phenomena in controlled environments. 
Beams of antiprotons together with a multi-purpose and compact 
detection system will provide unique tools to perform studies of 
the strong interaction. This will be achieved via precision 
spectroscopy of charmonium and open-charm states, an extensive 
search for exotic objects such as glueballs and hybrids, 
in-medium and hypernuclei spectroscopy, and more. An overview is 
given of the physics program of the ${\bar{\rm P}}$ANDA collaboration.}

\FullConference{European Physical Society Europhysics Conference on High Energy Physics,
EPS-HEP 2009,\\
		 July 16 - 22 2009\\
		 Krakow, Poland}

\begin{document}

\section{Introduction}

The fundamental building blocks of QCD are the quarks which 
interact with each other by exchanging gluons. QCD is well 
understood at short-distance scales, much shorter than the 
size of a nucleon ($<$~10$^{-15}$~m). In this regime, the basic 
quark-gluon interaction is sufficiently weak. In fact, many processes 
at high energies can quantitatively be described by perturbative QCD.
Perturbation theory fails when the distance among quarks becomes 
comparable to the size of the nucleon. As a 
consequence of the strong coupling, we observe the relatively heavy 
mass of hadrons, such as protons and neutrons, which is two orders of 
magnitude larger than the sum of the masses of the individual quarks. 
This quantitatively yet-unexplained behavior is related to the 
effect of chiral symmetry breaking.

The physics program of the ${\bar{\rm P}}$ANDA (anti-Proton ANnihilation at DArmstadt)
collaboration will address various questions related to the strong 
interactions by employing a multi-purpose detector system~\cite{PANDA,tdr_magnets,tdr_emc} at the 
High Energy Storage Ring for anti-protons (HESR) of the upcoming Facility 
for Anti-proton and Ion Research (FAIR)~\cite{FAIR}. The ${\bar{\rm P}}$ANDA 
collaboration aims to connect the perturbative and the non-perturbative QCD regions, 
thereby providing insight in the mechanisms of mass generation and 
confinement. For this purpose, a large part of the program will be 
devoted to charmonium spectroscopy; gluonic excitations, e.g. hybrids and glueballs;
open and hidden charm in nuclei. In addition, various other physics topics will be 
studied with ${\bar{\rm P}}$ANDA such as the hyperon-nucleon and hyperon-hyperon 
interactions via $\gamma$-ray spectroscopy of hypernuclei; CP violation studies 
exploiting rare decays in the D and/or $\Lambda$ sectors; studies of the structure 
of the proton by measuring Generalized Parton Distributions (Drell-Yan and 
Virtual-Compton Scattering), "spin" structure functions using polarized anti-protons, 
and electro-magnetic form factors in the time-like region.

\section{The experimental facility}

The key ingredient for the ${\bar{\rm P}}$ANDA physics program is a high-intensity and 
a high-resolution beam of antiprotons in the momentum range of 1.5 to 15~GeV/c.
Such a beam gives access to a center-of-mass energy range from 2.2 to 5.5~GeV/c$^2$ 
in $\bar p$$p$ annihilations. In this range, a rich spectrum of hadrons with
various quark configurations can be studied. In particular, hadronic states which contain 
charmed quarks and gluon-rich matter become experimentally accessible. 

The ${\bar{\rm P}}$ANDA detector will be installed at the High Energy Storage Ring, HESR,
at the future Facility for Antiproton and Ion Research, FAIR.  
FAIR provides a storage ring for beams of phase-space cooled antiprotons 
with unprecedented quality and intensity~\cite{HESR}. 
Antiprotons will be transferred to the HESR where internal-target experiments in 
the beam momentum range of 1.5 -- 15 GeV/c can be performed. Electron and stochastic 
phase space cooling will be available to allow for experiments with either 
high momentum resolution of about $\sim 10^{-5}$ at reduced luminosity or with 
high luminosity up to $2\times$10$^{32}$~cm$^{-1}$s$^{-1}$ with an enlarged momentum 
spread of $\sim 10^{-4}$.

The ${\bar{\rm P}}$ANDA detector is designed as a large acceptance multi-purpose setup. The 
experiment will use internal targets. It is conceived to use either pellets 
of frozen H$_2$ or cluster jet targets for the $\bar pp$ reactions, 
and wire targets for the $\bar pA$ reactions. 

To address the different physics topics, 
the detector needs to cope with a variety of final states and a large range of 
particle momenta and emission angles. 
At present, the detector is being designed to handle high rates of 10$^7$~annihilations/s , 
with good particle identification and momentum resolution for 
$\gamma,~ e,~ \mu,~ \pi,~ K$, and $p$ with the ability to measure 
$D,~ K^0_S$, and $\Lambda$ which decay at displaced vertices.  
Furthermore, the detector will have an almost 4$\pi$ detection coverage both for 
charged particles and photons. This is an essential requirement for an unambiguous partial 
wave analysis of resonance states. Various design and physics performance studies~\cite{ern09} 
are ongoing, partly making use of a dedicated computing framework for simulations and data 
analysis. 

\section{${\bar{\rm P}}$ANDA physics topics}

The level scheme of lower-lying bound $\bar c$$c$ states, charmonium, 
is very similar to that of positronium. These charmonium states
can be described fairly well in terms of heavy-quark potential models.
Precision measurements of the mass and width of the charmonium spectrum
give, therefore, access to the confinement potential in QCD.  
Extensive measurements of the masses and widths of the 1$^-$ $\Psi$ states 
have been performed at $e^+$$e^-$ machines where they can be formed directly
via a virtual-photon exchange. Other states, which do not carry the
same quantum number as the photon, cannot be populated directly, but only
via indirect production mechanisms. This is in contrast to the $\bar p$$p$ 
reaction, which can form directly excited charmonium states of 
all quantum numbers. As a result, the resolution in the mass and 
width of charmonium states is determined by the precision of the phase-space 
cooled beam momentum distribution and not by the (significantly poorer) 
detector resolution. The need for such a tool becomes evident by 
reviewing the many open questions in the charmonium sector.
For example, our understanding of the states above the $D$$\bar D$ threshold 
is very poor and needs to be explored in more detail. Recent experimental
evidences (see review~\cite{Bar06}) hint at a whole series of surprisingly 
narrow states with masses and properties which, so-far, cannot be interpreted 
consistently by theory. Besides the spectroscopy of charmonium states, 
${\bar{\rm P}}$ANDA will also provide
the capability to perform open-charm spectroscopy as the analog of
the hydrogen atom in QED (heavy-light system). Striking discrepancies
of recently discovered $D_{sJ}$ states by BaBar~\cite{Aub03} and CLEO~\cite{Bes03} 
with model calculations have been observed. Precision measurements of the
masses and widths of these states using antiprotons and by performing 
near-threshold scans are needed to shed light on these open problems.

The self-coupling of gluons in strong QCD has an important consequence, 
namely that QCD predicts hadronic systems consisting of only gluons, 
glueballs, or bound systems of quark-antiquark pairs with a strong 
gluon component, hybrids. These systems cannot be categorized as 
"ordinary" hadrons containing valence $q\bar q$ or $qqq$. 
The additional degrees of freedom 
carried by gluons allow glueballs and hybrids to have spin-exotic 
quantum numbers, $J^{PC}$, that are forbidden for normal mesons 
and other fermion-antifermion systems. States with exotic quantum numbers
provide the best opportunity to distinguish between gluonic hadrons 
and $q\bar q$ states. Exotic states with conventional quantum 
numbers can be identified by measuring an overpopulation of the 
meson spectrum and by comparing properties, like masses, quantum numbers,
and decay channels, with - for instance - predictions from 
Lattice Quantum Chromodynamics (LQCD) calculations.
The most promising energy range to discover unambiguously 
hybrid states and glueballs is in the region of 3-5 GeV/c$^2$, 
in which narrow states are expected to be superimposed on a 
structureless continuum. In this region, LQCD predicts an exotic 1$^{-+}$ 
$\bar cc$-hybrid state with a mass of 4.2-4.5~GeV/c$^2$ and a glueball
state around 4.5~GeV/c$^2$ with an exotic quantum number 
of $J^{PC}$=0$^{+-}$~\cite{Mor99,Bal04}. The $\bar pp$ production cross section of 
these exotic states are similar to conventional states and in the
order of 100~pb. All other states with ordinary quantum numbers are
expected to have cross sections of about 1~$\mu$b.

One of the challenges in nuclear physics is to study
the properties of hadrons and the modification of these 
properties when the hadron is embedded in a nuclear 
many-body system. Only recently it became experimentally evident that 
the properties of mesons, such as masses of $\pi$, $K$, and $\omega$ mesons, 
change in a dense environment~\cite{Suz04,Bar97,Lau99,Trn05}.
The ${\bar{\rm P}}$ANDA experiment provides a unique possibility to extend 
these studies towards the heavy-quark sector by exploiting the $\bar p$$A$ reaction.
For instance, an in-medium modification of the mass of the $D$ meson would imply a modification of the
energy threshold for the production of $D$ mesons, compared to a free mass.
In addition, a lowering of the $D$-meson mass could cause
charmonium states which lie just below the $D\bar D$ threshold for the $\bar p p$ 
channel to reside above the threshold for the $\bar p A$ reaction. In such a case,
the width of the charmonium state will drastically increase, which can experimentally
be verified. Although this is intuitively a simple picture, in practice the situation
is more complicated since the mass of various charmonium states might also change
inside the nuclear medium. 
Besides the indirect in-medium studies as described above, ${\bar{\rm P}}$ANDA will 
be capable to directly measure the in-medium spectral shape of charmonium states. 
This can be achieved by measuring the invariant mass of the di-lepton decay products. 
For the $\Psi(3770)$, for instance, models predict mass shifts of the order 
of --100~MeV~\cite{Lee04}, which are experimentally feasible to observe.

So far, this paper has concentrated on only a few of the topics which will be addressed
by the ${\bar{\rm P}}$ANDA collaboration. There exists, however, a large variety of other physics
topics which can ideally be studied with the ${\bar{\rm P}}$ANDA setup at the antiproton facility at FAIR. 
For example, there is growing interest within the ${\bar{\rm P}}$ANDA collaboration to make use of 
electro-magnetic probes, photons and leptons, in antiproton-proton annihilation. 
These probes will be used to study the structure of the proton by measuring 
Generalized Parton Distributions (GPDs), to determine quark distribution functions via Drell-Yan processes, 
and to obtain time-like electro-magnetic form factors by exploiting 
the $\bar pp\rightarrow e^+e^-$ reaction with an intermediate massive virtual photon. 
Furthermore, the ${\bar{\rm P}}$ANDA collaboration has the ambition to
perform hypernuclei experiments, which enables a study of the hyperon-nucleon and
hyperon-hyperon interactions. Finally, plans for symmetry violation 
experiments with ${\bar{\rm P}}$ANDA will open a window onto physics beyond the 
Standard Model of particle physics.


\section{Summary}

The ${\bar{\rm P}}$ANDA experiment at FAIR will address a wide range of
topics in the field of QCD, of which only a small
part could be presented in this paper. The physics program
will be conducted by using beams of antiprotons together with 
a multi-purpose detection system, which enables
experiments with high luminosities and precision resolution.
This combination provides unique possibilities to study 
hadron matter via precision spectroscopy of the 
charmonium system and the discovery of new hadronic matter, 
such as charmed hybrids or glueballs, as well as by measuring the properties
of hadronic particles in dense environments. New insights in the structure 
of the proton will be obtained by exploiting electromagnetic probes. 
Furthermore, the next generation of hypernuclei spectroscopy will be conducted 
by the ${\bar{\rm P}}$ANDA collaboration and at the J-PARC facility 
in Japan. 
To summarize, ${\bar{\rm P}}$ANDA has the ambition to 
provide valuable and new insights in the field of hadron physics which would bridge 
our present knowledge obtained in the field of perturbative QCD 
with that of non-perturbative QCD and nuclear structure.

\section*{Acknowledgments}

The author acknowledges the financial support from the University of 
Groningen and the GSI Helmholtzzentrum f\"ur Schwerionenforschung GmbH, Darmstadt.

\end{document}